\documentclass[pdflatex,sn-mathphys-ay]{sn-jnl}

\usepackage{graphicx}%
\usepackage{multirow}%
\usepackage{amsmath,amssymb,amsfonts}%
\usepackage{amsthm}%
\usepackage{mathrsfs}%
\usepackage[title]{appendix}%
\usepackage{xcolor}%
\usepackage{textcomp}%
\usepackage{manyfoot}%
\usepackage{booktabs}%
\usepackage{algorithm}%
\usepackage{algorithmicx}%
\usepackage{algpseudocode}%
\usepackage{listings}%
\usepackage{subcaption}
\usepackage{tikz}

\graphicspath{{img/}{../img/}}

\theoremstyle{thmstyleone}%
\theoremstyle{thmstyletwo}%

\theoremstyle{thmstylethree}%

\begin{document}

\title[Bayesian Modelling of Nonstationary Extreme Values Using a Nonparametric Hawkes Process]{ Bayesian Modelling of Nonstationary Extreme Values Using a Nonparametric Hawkes Process }


\author*[1]{\fnm{Gordon J. } \sur{Ross}}\email{gordon.ross@ed.ac.uk}

\author[2]{\fnm{Dean} \sur{Markwick}}\email{dean.markwick@talk21.com}

\affil*[1]{\orgdiv{Department of Mathematics}, \orgname{University of Edinburgh}, \orgaddress{ \country{UK}}}

\affil[2]{\orgdiv{Department of Statistics}, \orgname{University College London}, \orgaddress{ \country{UK}}}

\abstract{
Modelling and forecasting the occurrence of extreme events is especially difficult when the event process is nonstationary, with changes in both the rate at which extremes occur and the magnitude of the extremes when they occur. We approach this task by developing a Bayesian point process model for extreme events, which uses a self-exciting Hawkes process to model the rate at which extremes occur. The Hawkes process has a structure which allows events to occur in clusters, making it realistic for many types of data. We use a flexible Bayesian nonparametric approach based on the Dirichlet process to learn the temporal excitation pattern from the data. Further, we build on Extreme Value Theory by using a Generalised Pareto Distribution (GPD) to model the magnitudes of the extremes, with a hierarchical mark model allowing these magnitudes to vary across Hawkes-induced clusters. A hierarchical specification of the model results in partial pooling, allowing for more accurate GPD estimation even in clusters with only a small number of observations. We develop an MCMC algorithm to sample from the resulting hierarchical model. A simulation study confirms that the two flexible components improve prediction when the corresponding features are present in the data-generating mechanism, and across four real data sets the nonparametric Hawkes model with hierarchical GPD marks gives the best held-out predictive performance among the model variants considered.}




\maketitle

\section{Introduction}
Effective risk management often requires an estimate of the probability that large events will occur during a given period of time. For example, suppose that $m$ terrorist attacks have previously occurred over a period of $T$ years, at times $t_1, t_2,\ldots,t_m$. For each attack at time $t_i$, let $r_i$ be a mark denoting the corresponding number of fatalities. Based on this historical data, it may be desirable to produce a probabilistic estimate for the probability of another large attack occurring within some future time window \citep{porter_michael_self-exciting_2012}. Similar problems are also often considered in fields such as natural hazards modelling where the events correspond to earthquakes and the marks correspond to earthquake magnitudes 
\citep{bray_assessment_2013}, and in finance where the events are the times at which large changes in a company's stock price are observed, and the marks represent the size of the change \citep{kiriliouk_peaks_2019, chavez-demoulin_high-frequency_2012}.

 \begin{figure*}[t]
   \centering
   \begin{subfigure}[t]{0.48\textwidth}
     \centering
     \includegraphics[width=1\textwidth]{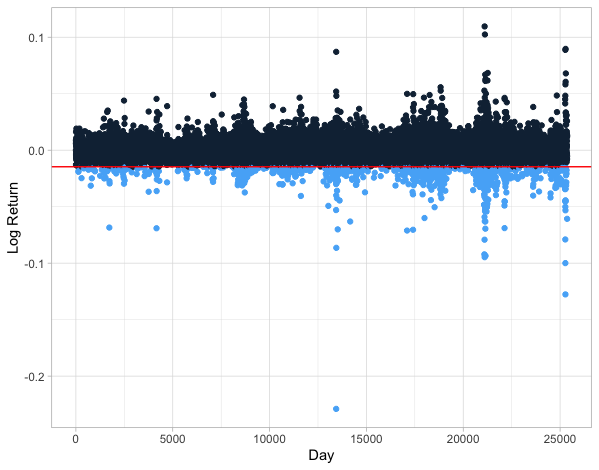}
     \caption{Daily log returns of the S{\&}P 500 index from 1951--2020.}
   \end{subfigure}\hfill
    \begin{subfigure}[t]{0.48\textwidth}
     \centering
     \includegraphics[width=1\textwidth]{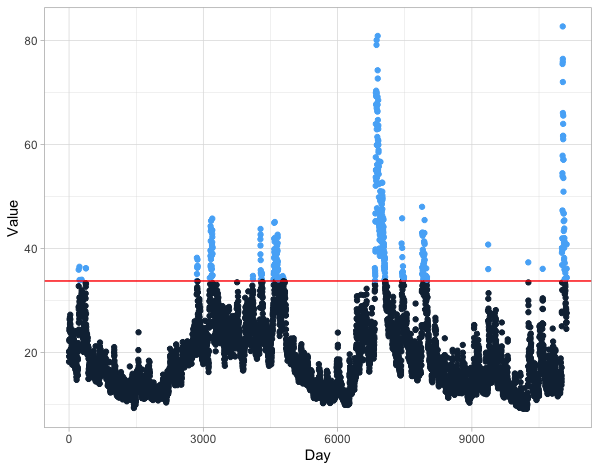}
     \caption{Daily closing values of the VIX index from 1990--2020.}
   \end{subfigure}
   \caption{Illustration of two stock market indexes with time variation in both the rate at which extreme values occur, and their magnitudes when they do occur. Extremes are defined as values above the $95^{th}$ percentile of the distribution (negative for the $S\&$P 500) and colored in light blue.}
   \label{fig:examplegraphs}
 \end{figure*}

This prediction task can often be broken down into two parts. First, a prediction is made for the number of large events which are likely to occur in a given period along with their occurrence times. Second, for each predicted event $r_i$, the probability $p(r_i > z)$ that the mark exceeds a threshold $z$ is then estimated. Direct estimation of this latter quantity is difficult since it usually involves inference about extreme quantiles of the distribution governing $r_i$. This can be highly sensitive to the parametric assumptions made about the distribution, with even small misspecification giving very inaccurate results. As such, it is usual to instead rely on asymptotic results from the field of Extreme Value Theory (EVT) to avoid the need to make strong parametric assumptions  \citep{coles_introduction_2001,davison_models_1990}.

For this purpose, the well-known Pickands--Balkema--de Haan (PBH) theorem from EVT states that given a sufficiently large threshold $u$, the conditional excess distribution $p(r_i-u \leq z \mid r_i > u)$ can be approximated by a Generalised Pareto Distribution (GPD), as long as the distribution satisfies certain regularity conditions \citep{balkema_residual_1974,davison_models_1990}. This has led to the widely-used Peaks-Over-Threshold (POT) approach for modelling extreme values, where the occurrence times $t_i$ of extremes are commonly modelled using a Poisson process, with the corresponding excess magnitudes $r_i-u$ following a GPD \citep{chavez-demoulin_estimating_2005}. However the direct application of this methodology to real-world problems is limited by the strong stationarity assumptions that it requires \citep{coles_introduction_2001}. Specifically, a straightforward EVT analysis is only possible if: a) the occurrence times $t_i$ of large values can be modelled by a possibly inhomogeneous Poisson process and b) the magnitudes exceeding the threshold are independent draws from a GPD with constant parameters.

Unfortunately these assumptions often do not hold. Figure~\ref{fig:examplegraphs} shows the daily values of two major stock market indexes. Both indexes are designed to give an easily calculable measure of the US stock market and an indicator of the general outlook and volatility of the economy. Many risk measures are based on the S{\&}P 500 daily closing price, and it is therefore of interest to predict when the next large drop could occur and how severe it could be \citep{chavez-demoulin_estimating_2005}. In Figure~\ref{fig:examplegraphs} we can see notable clustering of the extremes, where they tend to occur in bursts. Furthermore, there appears to be structural change in the distribution $p(r_i > z)$ of the extremes, with losses being structurally larger during certain time periods as well as more frequent. This is partly due to the well-known phenomenon of heteroskedasticity in stock returns as discussed by \citet{poon_modelling_2003}. It would hence be unwise to fit a single stationary GPD to these data.


Numerous extensions of EVT have been proposed to model non-stationarity in extreme events. An influential early paper suggested a parametric regression framework for the GPD parameters to allow variation over time \citep{davison_models_1990} and this idea has been extended in several ways \citep{northrop_threshold_2011, chavez-demoulin_estimating_2005}. Related work has proposed various more sophisticated models for the point process governing the occurrence of extremes, for example \citep{gyarmati-szabo_statistical_2011,Stindl2023IntradayRisk}, while other approaches involve an initial pre-processing step aimed at removing nonstationarity \citep{eastoe_modelling_2009}. A partial review of the extensive literature on EVT can be found in \cite{coles_introduction_2001}.

Building on the marked point process representation of threshold exceedances, \cite{kottas_bayesian_2007} and \cite{kottas_spatial_2012} developed Bayesian
nonparametric models for the associated intensity function. The flexibility of nonparametric estimation allows nonstationarity in both the time and mark domains to be handled naturally. However while their framework is well-suited to modelling historical data, it is less useful for making predictions about the occurrence of extremes in the future. This is because their point process representation effectively  smooths out the historical data rather than explicitly modelling the conditional intensity function of the point process, which makes it difficult to make predictions based on recent process behaviour.

This article develops a Bayesian marked Hawkes POT model for predicting future extreme events in situations where the exceedances are non-stationary in both the time and mark domains. The occurrence times of exceedances are modelled using a self-exciting Hawkes process \citep{hawkes_spectra_1971}, so that recent exceedances can increase the short-term probability of further extremes and induce clusters of events. Rather than imposing a fixed parametric form on the excitation mechanism, we estimate the Hawkes triggering kernel using a flexible Bayesian nonparametric model based on the Dirichlet process. We use the latent branching representation of the Hawkes process not only for posterior computation, but also as a way of linking temporal clustering with variation in the mark distribution. Specifically, exceedances belonging to different Hawkes-induced clusters are allowed to have different mark distributions, with a hierarchical prior used to borrow strength across clusters. This links temporal clustering in the exceedance process to variation in the magnitudes of the exceedances, while avoiding the need to specify a parametric time-varying model for the GPD parameters.

The main contributions of the paper are threefold. First, we develop a marked Hawkes POT model for forecasting future threshold exceedances, rather than only smoothing historical extremes. Second, we use a Dirichlet process mixture prior for the Hawkes triggering kernel, allowing the temporal excitation pattern to be learned flexibly from the data. Third, we link the latent branching structure of the Hawkes process to a hierarchical GPD mark model, allowing the scale of exceedance magnitudes to vary across Hawkes-induced regimes while sharing information across clusters. The resulting model is evaluated using held-out predictive scores for both exceedance times and magnitudes.

We begin Section~\ref{sec:evt} by reviewing the traditional methods of EVT for estimating $p(r > z)$ when the event process is stationary. We then continue in Section~\ref{sec:Methods} by introducing the Hawkes process and show how it can be used in a nonparametric manner which is suitable for when there is no strong theoretical motivation for particular parametric assumptions. In Section~\ref{subsec:Posterior} we detail a Metropolis--Hastings algorithm for sampling the full posterior distribution of our model parameters. Section~\ref{sec:Simulation} evaluates the behaviour of the method in a controlled simulation study, before Section~\ref{sec:Applications} applies the methodology to real-world data sets.

\section{Extreme Value Theory}
\label{sec:evt}

We first recall the standard peaks-over-threshold construction in the idealised stationary case. Suppose that $r_1,\ldots,r_m \sim F$ are a sequence of independent and identically distributed observations, and that interest lies in the probability of large values occurring. If the functional form of $F$ is known, then this can be computed directly from its quantile function after any unknown parameters have been estimated. However the functional form of $F$ is usually unknown, and a particular parametric form will have to be chosen based on both the observed data and theoretical considerations. Unfortunately, inference for extreme quantiles of $F$ is known to be highly sensitive to these parametric assumptions \citep{porter_michael_self-exciting_2012}.

To avoid specifying a parametric model for the full distribution $F$, it is common to instead use the peaks-over-threshold (POT) approach, which models only the distribution of the excess $r_i-u$ conditional on $r_i>u$, where $u$ is a threshold parameter \citep{coles_introduction_2001}. This approach is justified by the Pickands--Balkema--de Haan theorem \citep{balkema_residual_1974}, which states that for a sufficiently large threshold $u$, the conditional excess distribution can be approximated by a Generalised Pareto Distribution:

\textbf{Pickands--Balkema--de Haan (PBH) Theorem}: Suppose $r_1,\ldots,r_m$ are i.i.d. with distribution $F$. Let $F_u(z)=p(r_i-u\leq z\mid r_i>u)$ denote the conditional excess distribution function which describes the behaviour of $F$ above a given threshold $u$. Then, assuming that $F$ satisfies suitable regularity conditions, $F_u$ converges to the Generalised Pareto Distribution (GPD), i.e. $F_u(z)\rightarrow G(z\mid\sigma,\xi)$ as $u\rightarrow\infty$, where
\begin{equation*}
G(z\mid\sigma,\xi)=
\begin{cases}
1-\left(1+\xi z/\sigma\right)^{-1/\xi}_{+}, & \xi\neq 0,\\
1-\exp(-z/\sigma), & \xi=0,
\end{cases}
\end{equation*}
with $z>0$, scale parameter $\sigma>0$, shape parameter $\xi$, and $(a)_+=\max(a,0)$. Assuming these regularity conditions are satisfied, the POT approach to extreme value estimation is to choose a threshold $u$ sufficiently large to make the GPD a good approximation above $u$, estimate the GPD parameters $(\sigma,\xi)$, and then approximate $p(r_i>z\mid r_i>u)$ for $z>u$ by $1-G(z-u\mid\sigma,\xi)$.

The POT construction also has a natural marked point process interpretation \citep{coles_introduction_2001}. Write the original data as ordered pairs $(t_i,r_i)$ for $i=1,2,\ldots,m$, where $t_i$ denotes the time at which $r_i$ is observed. After deleting the pairs where $r_i<u$, suppose that $n$ exceedances remain, and write these as $(t_i,y_i)$ for $i=1,\ldots,n$, where $y_i=r_i-u$ denotes the excess over the threshold. In the simplest stationary formulation, the exceedance times $t_1,\ldots,t_n$ are governed by a homogeneous Poisson process, while the excess magnitudes $y_i$ are independent draws from a common $\operatorname{GPD}(\sigma,\xi)$ distribution. Equivalently, the pairs $\{(t_i,y_i)\}$ can be viewed as observations from a homogeneous marked point process whose time component is Poisson and whose mark distribution is GPD.

In this simplest formulation, the process is stationary in time: exceedances are equally likely to occur at any point in the observation window, and the excess magnitudes are independent draws from a common GPD. The model developed below relaxes both parts of this classical marked point process representation: the homogeneous Poisson process for exceedance times is replaced by a Hawkes process, and the common GPD mark distribution is replaced by a hierarchical model for clustered exceedance magnitudes.

\section{Nonstationarity of the Exceedance Process} 
\label{sec:Methods}
As above, let  $y_1,\ldots,y_n$ denote the values of the marks which exceed some threshold $u$. A direct application of either the PBH theorem or the above point process representation for predicting future extreme values requires these $y_i$ marks to be treated as independent and identically distributed, with their time occurrence following a homogeneous Poisson process. This is problematic for two reasons:

\begin{enumerate}
\item The point process governing the times at which the exceedances occur can be non-stationary. This was previously seen in  Figure \ref{fig:examplegraphs} where the exceedances fall into clusters, with no exceedances occurring for long periods of time followed by many occurring close together. 

\item The distribution $p(y_t)$ of the exceedances may also change over time. Again this was seen in Figure \ref{fig:examplegraphs} where the magnitude of the extremes tends to be higher in some time periods compared to others.
\end{enumerate}

A substantial amount of existing literature on extreme value theory relaxes these assumptions in various ways.
 For example in \citep{leadbetter_weak_1976}, it is shown that as long as exceedances satisfy some mixing conditions, they are no longer required to be independent and instead can display some local dependence. This local dependence is expressed through the extremal index $\theta$ \citep{hsing_exceedance_1988} and can be interpreted as the average clustering effect in the extreme values.  Similarly, much of the existing literature relies on specifying parametric models for the time-evolution of both the occurrence time process, and the distribution $p(y_t)$.  Although this is a reasonable approach for modelling historical non-stationarity, it typically does not allow for inference of the \textbf{conditional} exceedance distribution  $p(y_t > z \mid y_t > u, y_{1:(t-1)})$ which will often be the main object of interest when making predictions about the future. In many applications it will be important to assess the probability of an extreme value occurring at some particular time point $t$ (e.g. ``next week") which requires taking into account the recent history of the process.

We hence take a different approach which focuses on modelling the conditional exceedance process directly. This is based on the point process representation for the marked exceedance process $(t_i,y_i)$. However, rather than treating this process as homogeneous in the time-domain with a constant mark distribution, we instead use a representation that allows for conditional nonstationarity in both domains. Specifically, we write the marked point process in terms of a conditional occurrence intensity and a conditional mark density,
\begin{equation*}
  \lambda(t,y\mid H_t)=\lambda(t\mid H_t)f(y\mid t,H_t),
\end{equation*}
where $H_t$ denotes the history of the process up to time $t$. This allows for both types of non-stationarity to be incorporated. First, the conditional intensity $\lambda(t\mid H_t)$ controls the times at which the extremes occur, and can incorporate features such as temporal clustering. Second, the conditional mark density $f(y\mid t,H_t)$ controls the magnitude of the extremes when they occur, and may also vary with the history of the process. We will model $\lambda(t\mid H_t)$ as a Hawkes process, which allows non-stationarity and clustered behaviour to arise directly from the conditional intensity function, and model $f(y\mid t,H_t)$ using a hierarchical GPD-based mark model which allows the distribution of exceedance magnitudes to vary across clusters.

\begin{figure*}[t]
\centering
\begin{subfigure}[b]{0.45\textwidth}
\centering
\includegraphics[width=1\textwidth]{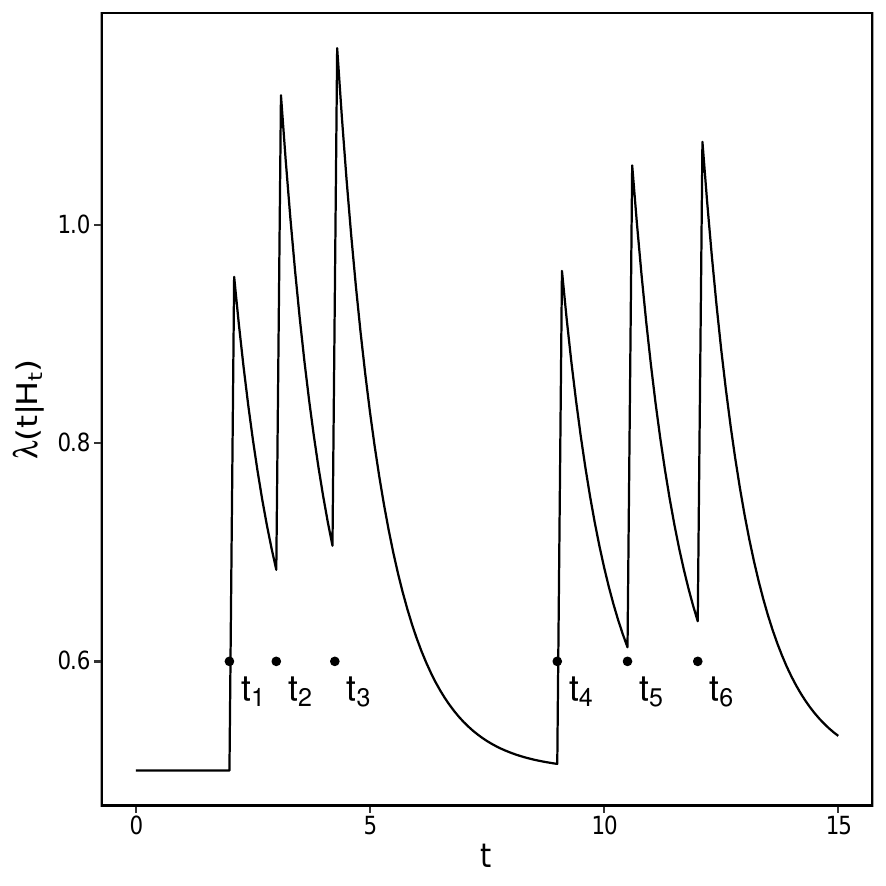}
\caption{Intensity function of a Hawkes process with events shown in Figure \ref{fig:HawkesPicture}}
\label{fig:HawkesPictureIntensity}
\end{subfigure}
\hfill
\begin{subfigure}[b]{0.45\textwidth}
\centering
\begin{tikzpicture}
\draw[ultra thick, black] (0,0+2) -- (5, 0+2);
\filldraw[black]  (1,0+2) circle (2.5pt);
\node[above] at (1,0+2) {$t_1$};
\draw[ultra thick, black] (1, 0+2) -- (1, -0.5+2);
\draw[ultra thick, black] (1,-0.5+2) -- (2.5, -0.5+2);
\filldraw[black] (3,0+2) circle (2.5pt);
\node[above] at (3,0+2) {$t_4$};
\draw[ultra thick, black] (3, 0+2) -- (3, -0.5+2);
\draw[ultra thick, black] (3,-0.5+2) -- (5, -0.5+2);
\filldraw[black] (1.5, -0.5+2) circle (2.5pt);
\node[below] at (1.5, -0.5+2) {$t_2$};
\filldraw[black] (2, -0.5+2) circle (2.5pt);
\node[below] at (2, -0.5+2) {$t_3$};
\filldraw[black] (4, -0.5+2) circle (2.5pt);
\node[above] at (4, -0.5+2) {$t_5$};
\draw[ultra thick, black] (4, -0.5+2) -- (4, -1+2);
\draw[ultra thick, black] (4, -1+2) -- (5, -1+2);
\filldraw[black] (4.5, -1+2) circle (2.5pt);
\node[above] at (4.5, -1+2) {$t_6$};
\end{tikzpicture}
\caption[Illustration of a Hawkes process]{Graphical representation of the structure of events arising in a simulated Hawkes process. Each black circle represents an event and shows how parent and child events form.}
\label{fig:HawkesPicture}
\end{subfigure}

\caption{Illustration of the branching structure and intensity function of a Hawkes process.}
\end{figure*}
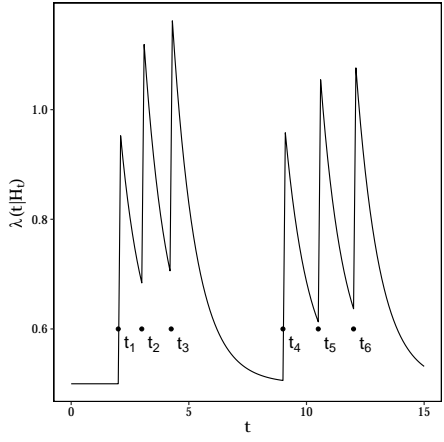
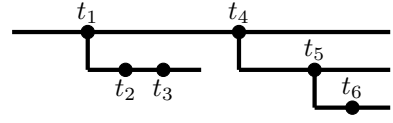

\subsection{Hawkes Process} \label{subsec:hawkes}
A point process on the interval $[0,T]$ can be characterised by its conditional intensity function $\lambda (t \mid H_t )$, where $H_t$ is the history of the process up to time $t$. This conditional intensity function satisfies the counting property of a point process:
\begin{equation*}
\lambda (t \mid H_t) = \lim _{\Delta t \rightarrow 0} \frac{\text{Pr} (N(t + \Delta t) - N(t) = 1 \mid H_t )}{\Delta t},
\end{equation*}
where $N(t)$ is the number of events that occur in the interval $\left[ 0, t \right]$  \citep{daley_introduction_2003}. The conditional intensity function is dependent on the history of the process so that past events can influence the rate at which future events occur. The Hawkes process \citep{hawkes_spectra_1971,hawkes_cluster_1974} is a type of self-exciting point process, with conditional intensity:
\begin{equation}
\lambda (t \mid H_t) = \mu (t) +  \sum _{t_i < t} \kappa h(t - t_i),
\label{eq:HawkesIntensity}
\end{equation}
where $t_i$ denotes the time at which the $i$th event occurred, $\mu (t) > 0$ is a background intensity function, $\kappa > 0$ is a constant and $h(\cdot)$ is a probability density that integrates to 1, known as the triggering or excitation kernel.  For ease of exposition we assume $\mu(t) = \mu$ is constant although time-variation in $\mu$ can be accounted for using a similar approach to \citep{deutsch2025cannibalisation}.  The essence of the Hawkes process is that the occurrence of an event at time $t$ makes it more likely for further events to occur soon afterwards, since the intensity increases by an amount controlled by the excitation kernel $\kappa h(\cdot)$. This leads to events occurring in clusters, making the Hawkes process well suited to modelling non-stationary exceedances.  As an illustration, Figure \ref{fig:HawkesPictureIntensity} plots the  conditional intensity function of a sample Hawkes process, where it can be seen that each event increases the conditional intensity, leading to further events occurring and thus events arriving in clusters.

The Hawkes process can also be interpreted as a branching process as first noted by \citep{hawkes_cluster_1974}. At each time point $t$, suppose that $n_t$ previous events have occurred. Then, the intensity function in Equation \ref{eq:HawkesIntensity} can be interpreted as a linear superposition of $n_t+1$ Poisson processes where the first has intensity function $\mu(t)$ and the other $n_t$ triggered processes each have intensity function $\kappa h(t-t_j)$ for each previous event $t_j$. Under this interpretation, an event which occurs at time $t_i$ will either have been generated by the background process $\mu (t)$ or by a triggered process, in which case we say that $t_i$ is a child of the event $t_j$ which triggered it. 

Figure \ref{fig:HawkesPicture} shows a sample realisation of the Hawkes process, to illustrate how a branching structure appears naturally. Events $t_1$ and $t_4$ are exogenous (i.e. have no parent) and were generated by the background intensity $\mu (t)$. Events $t_2, t_3, t_5$ were all triggered by events from the background process, whereas event $t_6$ was triggered by $t_5$. This leads us to introduce a latent variable for each event $\mathbf{B} = \{B_1, B_2, \ldots , B_n \}$ that describes the branching structure, where $B_i$ is the index of the event which caused $t_i$, with $B_i = 0$ if  $t_i$ is caused by the background process. Using Figure \ref{fig:HawkesPicture} as an example, the branching structure is: $\mathbf{B} = \{0, 1, 1, 0, 4, 5 \}$. This branching variable and its effect on the posterior distribution leads to an efficient computational method for sampling the parameters of the Hawkes process which we will discuss later. 


\subsubsection{A Nonparametric Kernel}
In most previous applications of the Hawkes process, the triggering kernel $h(t)$ has been specified parametrically and fitted using frequentist techniques \citep{porter_michael_self-exciting_2012, balderama_application_2012}. However in most realistic applications it will not be obvious which parametric form is most appropriate. As such, there has been recent interest in nonparametric specifications of the kernel function, typically within a frequentist framework using kernel density estimation or histogram estimators  \citep{zhuang_stochastic_2002,fox_spatially_2016,bacry_non-parametric_2012}. Additionally, \cite{markwick_d_and_ross_g_j_hierarchical_2020} presented a Bayesian version of the Hawkes process which uses a Dirichlet process mixture prior to nonparametrically model the background rate $\mu(t)$ within a Bayesian setting. We extend their work to form a nonparametric estimate of $h(t)$. 
 
The Dirichlet process (DP) is a stochastic process whose draws are distributions themselves, which is commonly used as a prior over the space of probability distributions. A DP is specified by its base distribution $G_0$ and a concentration parameter $\alpha_{DP}$. If $G$ is a draw from a DP then we write $G \sim DP (\alpha_{DP}, G_0)$. It can be shown \citep{ishwaran_gibbs_2001,sethuraman_constructive_1994} that $G$ can be written in stick-breaking form as:
\begin{align}
\begin{split}
G & = \sum _{i=1} ^{\infty} w _i \delta _{\psi _i}, \quad \psi _i  \sim G_0, \\
w_i & = z_i \prod _{j=1} ^{k-1} (1-z_j) , \quad z_j \sim \text{Beta} (1, \alpha_{DP}),
\label{eqn:stick}
\end{split}
\end{align}
where $\delta _{\psi _i}$ are point masses located at $\psi _i$ and the $z_j$'s are independent.
 From this, it can be seen that distributions drawn from a DP are almost surely discrete.  The DP is a conjugate prior in the following sense: if $G \sim DP (\alpha_{DP}, G_0)$ and $\mathbf{z} = (z_1,\ldots,z_n)$ are independent draws from $G$ then the posterior distribution for $G$ is also a DP $p(G|\mathbf{z}) = DP(\alpha', G_0')$ where \citep{ferguson_bayesian_1973,antoniak_mixtures_1974}:

\begin{align}
\begin{split}
\alpha' &= \alpha+n \\
G_0' &=  \frac{\alpha_{DP} }{\alpha_{DP} + n} G_0 + \frac{n}{\alpha_{DP} + n} \sum_{i=1} ^n  \frac{1}{n} \delta _{z_i}. 
\label{eqn:conjugate}
\end{split}
\end{align}

 Since our aim is to use the DP as a nonparametric prior on the excitation kernel $h(\cdot)$ of a continuous time process, the discreteness of DP samples is problematic. We hence follow a standard approach \citep{escobar_bayesian_1995,neal_markov_2000} and convolve $G$ with a continuous mixture kernel $k$. This produces a mixture model with an infinite number of components, leading to the specification:

\begin{align}
\begin{split}
h(t) & = \int k(t \mid \phi) \mathrm{d} G(\phi ), \\
\phi & \sim G, \\
G & \sim \text{DP} (\alpha_{DP}, G_0), 
\label{eqn:mixturespec}
\end{split}
\end{align}
where $G_0$ is the base distribution of the Dirichlet process and $\alpha_{DP}$ is the concentration parameter. By using an infinite mixture, the shape of $h(t)$  can be flexibly learned from the data in order to incorporate features such as heavy tails and multimodality that might be missed when using a parametric kernel. To aid computation, we choose a Lognormal distribution for the kernel $k(\cdot)$ so that $\phi = \{\mu,\sigma^2\}$. The Lognormal mixture model is a flexible choice which allows for potential multimodality and heavy tails in $h(\cdot)$. It also admits a conjugate prior which will simplify parameter inference. Specifically, when using the Lognormal kernel we choose the base measure to be $G_0 = N (\mu \mid \mu _0, \frac{\sigma ^2}{k_0}) \text{Inv-Gamma} (\sigma^2 \mid  \alpha _0, \beta _0)$ where the prior parameters are taken to be weakly informative: $\{ \mu_0=0, k_0=1, \alpha _0=1, \beta _0=1 \}$.


\subsection{Extreme Value Theory and the GPD}

The observations in an extreme value event process consist of pairs of values $(t_i,y_i)$, where $t_i$ is the time at which the $i$th extreme event occurs and $y_i=r_i-u$ is the excess magnitude above a given threshold $u$. We use the Hawkes process to model the occurrence times of the extreme events, i.e. the $t_i$ values. We now discuss our non-stationary model for the magnitudes $y_i$. In the simplest application of the PBH theorem we could assume that, for some threshold $u$, the observations are independent and identically distributed samples from a GPD,
\begin{equation*}
y_i\sim \operatorname{GPD}(\sigma,\xi),
\end{equation*}
with common parameters $\sigma$ and $\xi$. These parameters can then be estimated, which allows us to compute the future probability of large excesses. However, this assumes that the distribution of the $y_i$ values is constant over time, which is not the case in many real-world applications. In the traditional extreme value literature these problems are often addressed by allowing the parameters of the GPD to depend on time, and applying common regression estimation techniques \citep{coles_introduction_2001}. However, this requires choosing a particular parametric form for the time dependence, which can lead to under-fitting or over-fitting depending on the number of parameters chosen. It also does not easily allow for the formation of clusters in the data, which were clearly seen in Figure~\ref{fig:examplegraphs}. Instead, we propose a method that groups the data into local temporal regimes using the branching structure of the Hawkes process, and allows the mark distribution to vary across these regimes. A hierarchical model is then used to pool information across clusters so that the cluster-level quantities can be estimated accurately even when the number of observations in each cluster is small.

Recall from Section~\ref{subsec:hawkes} that the Hawkes process naturally produces clusters of events, represented by the branching variables $\mathbf{B}=(B_1,\ldots,B_n)$, where $B_i=j$ if $(t_i,y_i)$ was generated by the Poisson process spawned by the event that occurred at time $t_j$, and $B_i=0$ if $(t_i,y_i)$ was generated by the background process. The occurrence of events in the background process naturally splits the observations into clusters, since each background event can result in a cascade of further events, as shown in Figure~\ref{fig:HawkesPicture}. A natural alternative would be to define clusters genealogically, by grouping each background event with all of its descendants. However, when several background events occur close together, their descendants may be interleaved in time, which makes the resulting mark regimes less natural as local temporal regimes.

As such, we instead define each cluster as the temporal interval which elapses between each pair of background events. Let $a_1<a_2<\cdots<a_K$ denote the indices of the background events, so that $B_{a_k}=0$ for $k=1,\ldots,K$. Define $s_k=t_{a_k}$ for $k=1,\ldots,K$, and set $s_{K+1}=T$. The $k$th cluster is then
\begin{equation*}
C_k=\{y_i:s_k\leq t_i<s_{k+1}\},\qquad k=1,\ldots,K.
\end{equation*}
Since the first event is necessarily a background event, these sets form a partition of the observed exceedances. Referring back to Figure~\ref{fig:HawkesPicture}, this sequence would hence have two clusters, the first consisting of events $\{t_1,t_2,t_3\}$ and the second consisting of events $\{t_4,t_5,t_6\}$.

A fully separate GPD fit within each cluster would be unstable, since the Hawkes branching structure can produce many clusters containing only a small number of exceedances. We therefore use the clustering to model local variation in the scale of exceedances, while estimating the tail-shape parameter globally. Conditional on the branching structure $\mathbf{B}$, if $y_i\in C_k$ we write
\begin{align}
\begin{split}
y_i\mid\mathbf{B},\sigma_k,\xi &\sim \operatorname{GPD}(\sigma_k,\xi),\\
\log\sigma_k &= \log\sigma_0+\tau_\sigma z_k,\qquad z_k\sim N(0,1),
\end{split}
\label{eqn:GPDcluster}
\end{align}
where $\sigma_k$ is the GPD scale in cluster $C_k$ and $\xi$ is a shared shape parameter. We note that this specification is conditional on the branching structure $\mathbf{B}$ which divides the sequence of observations into clusters. Since $\mathbf{B}$ is unknown, it must be estimated, which will be discussed in the next section. The marginal distribution of each $y_i$ can then be obtained by marginalising $\mathbf{B}$ out of the joint posterior distribution for all model parameters.

The hierarchical prior on the cluster scales allows information to be borrowed across clusters \citep{gelman_multilevel_2006,gelman_bayesian_2014}. We use weakly informative priors on the scale hierarchy and a regularising prior on the shared shape parameter:
\begin{align}
\begin{split}
\log\sigma_0 &\sim N(0,1),\\
\tau_\sigma &\sim N^+(0,0.5^2),\\
\xi &\sim N(0,0.2^2)\quad\text{truncated to }(-0.25,\infty),
\end{split}
\label{eq:gpdModel}
\end{align}
where $N^+$ denotes a Normal distribution truncated to the positive real line. The shared-shape specification is a deliberate regularisation: clusters may differ in exceedance scale, which captures changes in the typical magnitude of exceedances, while the tail-shape parameter is learned globally from all exceedances. The lower truncation on $\xi$ keeps inference away from the irregular finite-endpoint boundary of the GPD likelihood.

\section{Posterior Inference} \label{subsec:Posterior}

Posterior inference for our model involves estimating the Hawkes process parameters $\Theta_{HP}=\{\mu,\kappa,h(\cdot),\mathbf{B}\}$, which govern the exceedance times $t_1,\ldots,t_n$, and the GPD mark parameters
\begin{equation*}
\Theta_{GPD}=\{z_1,\ldots,z_K,\log\sigma_0,\tau_\sigma,\xi\},
\end{equation*}
which govern the excess magnitudes $y_i$ through $\sigma_k=\exp(\log\sigma_0+\tau_\sigma z_k)$. Here $K$ is not fixed in advance, but is determined by the current branching structure $\mathbf{B}$. 

Our MCMC sampler uses a modular posterior factorisation in which the Hawkes process is learned from the exceedance times and the mark model is fitted conditional on posterior draws of the induced temporal clustering. This deliberately prevents the exceedance magnitudes from feeding back into the estimation of the temporal branching structure. The choice is motivated by the interpretation of the Hawkes process as a model for exceedance occurrence times, with the mark model used to describe variation in excess magnitudes across the resulting temporal regimes.  We write $Y=\{(t_1,y_1),\ldots,(t_n,y_n)\}$ for the observed exceedances, and discuss each block of the sampler in turn.

\subsection{Sampling for the Hawkes Process}

To sample the Hawkes process parameters we use a similar approach to \citep{ross2021bayesianETAS,rasmussen_bayesian_2013}, with an extension for the nonparametric Dirichlet process model for $h(\cdot)$. The latent branching variables $\mathbf{B}=(B_1,\ldots,B_n)$ decompose the Hawkes likelihood into approximately independent parameter blocks. Given this branching structure, the event indices can be partitioned into the sets $S_0,S_1,\ldots,S_n$, where
\begin{equation*}
S_j=\{i:B_i=j\}.
\end{equation*}
Thus $S_0$ is the set of all events that were created by the background process with intensity $\mu$, and each $S_j$ is the set of events that were produced by the process triggered by event $t_j$. Conditional on $\mathbf{B}$, the Hawkes likelihood for the event times can be written as \citep{rasmussen_bayesian_2013}
\begin{equation}
L_{HP}(\mu,\kappa,h;t,\mathbf{B})=
\mu^{|S_0|}e^{-\mu T}
\prod_{j=1}^{n}
\left\{
 e^{-\kappa H(T-t_j)}\kappa^{|S_j|}
 \prod_{i\in S_j}h(t_i-t_j)
\right\},
\label{eq:finallik}
\end{equation}
where $t=(t_1,\ldots,t_n)$, $H(z)=\int_0^z h(u)\,du$, and $|S_j|$ denotes the number of events in set $S_j$. The point of this latent variable parameterisation is that the posterior for $\mu$ is independent of the other Hawkes parameters, while the posterior dependence between $\kappa$ and $h(\cdot)$ is substantially reduced. The MCMC sampling for $\mathbf{B}$, $\mu$ and $\kappa$ then follows the same general scheme as in \citet{ross2021bayesianETAS}, before updating the nonparametric triggering kernel $h(\cdot)$.

\paragraph{Sampling from $p(\mathbf{B}\mid\Theta_{HP}^{(s-1)},Y)$:} At each time point $t$, the Hawkes intensity in Equation~\eqref{eq:HawkesIntensity} is a superposition of a background Poisson process with intensity $\mu$ and multiple triggered Poisson processes, one for each previous event. Conditional on the Hawkes parameters, the posterior allocation probabilities are proportional to the contribution of each component to the total intensity. Thus, for event $t_i$,
\begin{align}
\begin{split}
\Pr(B_i=0\mid\Theta_{HP}^{(s-1)},Y)
&=\frac{\mu}{\lambda(t_i\mid H_{t_i})},\\
\Pr(B_i=j\mid\Theta_{HP}^{(s-1)},Y)
&=\frac{\kappa h(t_i-t_j)}{\lambda(t_i\mid H_{t_i})},\qquad j=1,\ldots,i-1,
\end{split}
\label{eq:ParentProb}
\end{align}
where $\lambda(t_i\mid H_{t_i})=\mu+\sum_{j<i}\kappa h(t_i-t_j)$ is the total conditional intensity at $t_i$. Since these distributions are discrete, each $B_i$ can be directly sampled from its posterior using multinomial sampling. The first event is necessarily assigned to the background process.

\paragraph{Sampling from $p(\mu\mid\mathbf{B},Y)$:} Conditional on $\mathbf{B}$, the events in set $S_0$ follow a homogeneous Poisson process with intensity function $\mu$. We use a conjugate prior $p(\mu)=\operatorname{Gamma}(\alpha_\mu,\beta_\mu)$, where $\beta_\mu$ is a rate parameter, leading to the posterior distribution
\begin{equation*}
\mu\mid\mathbf{B},Y\sim \operatorname{Gamma}\left(\alpha_\mu+|S_0|,\beta_\mu+T\right),
\end{equation*}
which can be sampled from exactly.

\paragraph{Sampling from $p(\kappa\mid\mathbf{B},h,Y)$:} From Equation~\eqref{eq:finallik}, the conditional posterior for $\kappa$ is
\begin{equation*}
p(\kappa\mid\mathbf{B},h,Y)
\propto p(\kappa)\kappa^{\sum_{j=1}^{n}|S_j|}
\exp\left\{-\kappa\sum_{j=1}^{n}H(T-t_j)\right\}.
\end{equation*}
If $p(\kappa)=\operatorname{Gamma}(\alpha_\kappa,\beta_\kappa)$, this gives the conjugate update
\begin{equation*}
\kappa\mid\mathbf{B},h,Y\sim
\operatorname{Gamma}\left(\alpha_\kappa+\sum_{j=1}^{n}|S_j|,
\beta_\kappa+\sum_{j=1}^{n}H(T-t_j)\right),
\end{equation*}
again using the rate parametrisation. If the subcriticality condition $\kappa<1$ is enforced through the prior, the same update is used with this Gamma distribution truncated to the interval $(0,1)$.

\paragraph{Sampling from $p(h(\cdot)\mid\mathbf{B},\mu,\kappa,Y)$:} For each event $t_i$ that is not in the background process, let $x_i=t_i-t_{B_i}$ be the observed triggering lag. The likelihood contribution involving $h(\cdot)$ is
\begin{equation*}
\left\{\prod_{i:B_i>0}h(x_i)\right\}
\exp\left\{-\kappa\sum_{j=1}^{n}H(T-t_j)\right\}.
\end{equation*}
The first term is the usual density contribution from the observed triggering lags, while the second term is the integrated hazard contribution from the finite observation window. We update $h(\cdot)$ using the conjugate Dirichlet process mixture update for the observed lags, with a Metropolis--Hastings correction for the integrated hazard term.


The mixture model in Equation~\eqref{eqn:mixturespec} can be rewritten as
\begin{equation*}
x_i\sim k(x_i\mid\phi_i),\qquad
\phi_i\sim G,
\qquad G\sim \operatorname{DP}(\alpha_{DP},G_0).
\end{equation*}
We use the Chinese Restaurant Process sampler of \citet[Algorithm 4]{neal_markov_2000} to update each mixture parameter $\phi_i$ given the current triggering lags. Given these $\phi_i$ values, the posterior for $G$ is $p(G\mid\alpha_{DP},G_0,\{\phi_i\})$, which from the conjugacy property in Equation~\eqref{eqn:conjugate} is also a DP distribution. A proposal $G^\star$ can then be drawn from this posterior using truncated stick breaking \citep{ishwaran_gibbs_2001}. Specifically, for a large truncation level $L$,
\begin{equation*}
G^\star=\sum_{\ell=1}^{L}w^\star_\ell\delta_{\psi^\star_\ell},
\qquad \psi^\star_\ell\sim G'_0,
\end{equation*}
with weights
\begin{equation*}
w^\star_\ell=v^\star_\ell\prod_{r<\ell}(1-v^\star_r),
\qquad v^\star_\ell\sim \operatorname{Beta}(1,\alpha'_{DP}),
\end{equation*}
where $G'_0$ and $\alpha'_{DP}$ are the posterior values from Equation~\eqref{eqn:conjugate}. In our implementation we used $L=1000$. Given $G^\star$, the proposed triggering kernel $h^\star$ and its cumulative distribution function $H^\star$ are fully defined, with
\begin{equation*}
H^\star(z)=\sum_{\ell=1}^{L}w^\star_\ell\Phi(z\mid\psi^\star_\ell),
\end{equation*}
where $\Phi$ is the cumulative distribution function of the Lognormal mixture kernel $k$. Since this proposal is based on the posterior for the observed triggering lags, the Metropolis--Hastings acceptance probability only needs to correct for the integrated hazard term, giving
\begin{equation*}
\min\left\{1,
\exp\left[-\kappa\sum_{j=1}^{n}\{H^\star(T-t_j)-H(T-t_j)\}\right]
\right\}.
\end{equation*}

\subsection{Sampling for the GPD}

Given a branching structure $\mathbf{B}$, the excess magnitudes $y_1,\ldots,y_n$ are divided into clusters based on Equation~\eqref{eqn:GPDcluster}. Suppose there are $K$ such clusters under the current branching structure. Conditional on this partition, the posterior for the mark model is proportional to
\begin{equation*}
p(\log\sigma_0,\tau_\sigma,\xi)
\prod_{k=1}^{K}
\left\{
 p(z_k)
 \prod_{i:y_i\in C_k}g(y_i\mid\sigma_k,\xi)
\right\},
\end{equation*}
where $g(\cdot\mid\sigma_k,\xi)$ is the GPD density and
\begin{equation*}
\sigma_k=\exp(\log\sigma_0+\tau_\sigma z_k).
\end{equation*}
The GPD density is evaluated subject to its usual support constraint $1+\xi y_i/\sigma_k>0$. Conditional on the current clustering, we update the mark-model parameters using Hamiltonian Monte Carlo as implemented in Stan \citep{carpenter_stan_2017}. Since the branching structure, and hence the clusters, can change between MCMC iterations, the cluster-specific scales are treated as local to the current partition rather than being assigned persistent labels across all MCMC iterations. This avoids the need to match cluster labels between different branching structures.

\subsection{Posterior prediction}

The fitted model can be used to obtain posterior predictive distributions for future exceedances. For each posterior draw of the model parameters, we simulate the Hawkes process forward over a future window $(T,T+H]$ conditional on the observed history up to time $T$. This produces a draw of the future exceedance count
\begin{equation*}
N_H=N(T+H)-N(T),
\end{equation*}
along with the corresponding future exceedance times. Conditional on the simulated branching structure, excess magnitudes are then drawn from the appropriate GPD distribution. Events assigned to an existing cluster use the corresponding cluster-level scale parameter and the shared shape parameter. New background events initiate new clusters; for these clusters a new latent $z_{\mathrm{new}}\sim N(0,1)$ is drawn, giving
\begin{equation*}
\log\sigma_{\mathrm{new}}=\log\sigma_0+\tau_\sigma z_{\mathrm{new}},
\end{equation*}
while the same global shape parameter $\xi$ is used.

Repeating this procedure across posterior draws gives predictive distributions for quantities such as the number of exceedances in the future window, the maximum future excess
\begin{equation*}
M_H=\max\{Y_i:T<t_i\leq T+H\},
\end{equation*}
and tail probabilities of the form
\begin{equation*}
\Pr(M_H>z\mid Y),
\end{equation*}
for high levels $z$. These are the predictive quantities used in the empirical analysis below, in addition to the out-of-sample predictive likelihood. For numerical stability the GPD model is fitted to excesses divided by a training-set scale factor. Predictive densities reported on the original excess scale include the corresponding Jacobian correction.

\subsection{Computational considerations}

The most expensive part of the sampler is the update of the branching variables, since each event can in principle be assigned to any earlier event. A direct implementation of this step is therefore $O(n^2)$ in the number of exceedances. In the present setting this is typically manageable because the model is fitted only to threshold exceedances, so $n$ is much smaller than the length of the original time series. In larger applications the branching update can be accelerated by ignoring parent assignments whose triggering lag has negligible density under the current kernel. The truncation level $L$ in the stick-breaking representation is chosen sufficiently large that the remaining stick mass is negligible. In practice we monitor convergence using trace plots and posterior summaries for $\mu$, $\kappa$, the GPD scale-hierarchy parameters, and the main posterior predictive quantities.

Table~\ref{tab:prior-settings} summarises the prior distributions and fixed hyperparameter settings used in the fitted models. For each real-data analysis we ran four chains for both the Exponential-Hawkes and DP-Hawkes models, using 10,000 iterations per chain and discarding the first 2,000 iterations as burn-in. No thinning was used. The dirichletprocess R package \citep{ross2018dirichletprocess} was used to implement the above sampling of the DP. For the hierarchical GPD mark model, we used 100 evenly spaced retained Hawkes posterior draws per model as representative branchings. The final hierarchical GPD fits were run in Stan using four chains, 2,000 iterations per chain, 1,000 warm-up iterations, \texttt{adapt\_delta}=0.99, and \texttt{max\_treedepth}=15.

\begin{table}[t]
\centering
\begin{tabular}{lll}
\hline
Component & Parameter & Prior or setting \\
\hline
Hawkes background & $\mu$ & $\operatorname{Gamma}(0.1,0.1)$ \\
Hawkes branching & $\kappa$ & Uniform on $(0,1)$ \\
Exponential kernel & $\beta$ & Uniform on $(0,100)$ \\
DP concentration & $\alpha_{DP}$ & $\operatorname{Gamma}(2,4)$ \\
DP log-lag base measure & $G_0$ & Gaussian base measure with parameters $(0,1,1,1)$ \\
GPD global log-scale & $\log\sigma_0$ & $N(0,1)$ \\
GPD scale variation & $\tau_\sigma$ & half-$N(0,0.5^2)$ \\
GPD shape & $\xi$ & $N(0,0.2^2)$, truncated below at $-0.25$ \\
\hline
\end{tabular}
\caption{Prior distributions and fixed hyperparameter settings used in the fitted models. Gamma distributions use the shape/rate parameterisation. The DP mixture is fitted to log triggering lags; the Gaussian base-measure
parameters are \((0,1,1,1)\), and the DP concentration parameter uses the
\(\mathrm{Gamma}(2,4)\) prior.}
\label{tab:prior-settings}
\end{table}

\section{Simulation Study}\label{sec:Simulation}

We first conduct a simulation study to assess our models in a setting where the true data-generating mechanism is known. We consider a $2\times2$ design in which the true Hawkes triggering kernel is either Exponential or a two-component mixture, and the true mark distribution is either an iid GPD or a hierarchical GPD with clusters.

In all scenarios, events are generated on the interval $[0,1000]$ from a Hawkes process with background rate $\mu=0.10$ and branching parameter $\kappa=0.55$. The first 800 time units are used for training and the remaining 200 for testing and prediction. In the exponential-kernel scenarios the triggering density is $h(t)=\beta\exp(-\beta t)$ with $\beta=1$. In the mixture-kernel scenarios the triggering density is
\[
    h(t)=0.7\,\mathrm{Lognormal}(-0.3,0.35^2)
       +0.3\,\mathrm{Lognormal}(1.2,0.45^2),
\]
which produces a non-exponential excitation pattern with both short- and longer-lag triggering behaviour. For the mark distribution we set $\sigma_0=1$ and $\xi=0.15$. In the iid mark scenarios, all excesses are generated from a common $\mathrm{GPD}(\sigma_0,\xi)$ distribution. In the hierarchical mark scenarios, excesses in cluster $C_k$ are generated from $\mathrm{GPD}(\sigma_k,\xi)$, where
\[
    \log \sigma_k = \log \sigma_0 + \tau_\sigma z_k,
    \qquad z_k\sim N(0,1),
\]
with $\tau_\sigma=1$. As in the fitted model, clusters are defined as temporal intervals between background events in the Hawkes branching structure.

For each simulated data set we fit the following four models:

\begin{enumerate}
    \item Hawkes process with a parametric Exponential kernel $h(t)$, with the $y_t$ values treated as independent draws from a single $\operatorname{GPD}(\sigma,\xi)$ distribution.
    \item Hawkes process with a nonparametric Lognormal DP mixture kernel $h(t)$, with the $y_t$ values treated as independent draws from a single $\operatorname{GPD}(\sigma,\xi)$ distribution.
    \item Hawkes process with a parametric Exponential kernel $h(t)$, with the $y_t$ values modelled using the hierarchical cluster-scale GPD model described in Section~\ref{sec:Methods}.
    \item Hawkes process with a nonparametric Lognormal DP mixture kernel $h(t)$, with the $y_t$ values modelled using the hierarchical cluster-scale GPD model described in Section~\ref{sec:Methods}.
\end{enumerate}

Performance is measured using held-out log predictive scores on the test period. To make the comparisons stable across simulated data sets, Table~\ref{tab:simulation-score-deltas} reports score differences relative to the Exp+iid baseline (Model 1 above) within each replicate. Positive values therefore indicate an improvement over the baseline model. We ran 10 replicates for each scenario.


\begin{table}[t]
\centering
\begin{tabular}{lcccc}
\hline
Truth scenario & Exp+iid & DP+iid & Exp+hier. & DP+hier. \\
\hline
Exponential kernel, iid marks & 0.000 & 0.035 (0.176) & 0.104 (0.090) & 0.176 (0.159) \\
Exponential kernel, hier. marks & 0.000 & -0.176 (0.197) & 2.125 (0.935) & 2.322 (1.323) \\
Mixture kernel, iid marks & 0.000 & 1.703 (0.619) & -0.075 (0.061) & 1.665 (0.548) \\
Mixture kernel, hier. marks & 0.000 & 1.922 (0.445) & 0.806 (0.689) & 2.675 (0.871) \\
\hline
\end{tabular}
\caption{Mean held-out (test set) log predictive score differences in the simulation study, relative to the Exp+iid model within each replicate. Monte Carlo standard errors are shown in parentheses. Positive values indicate improvement over the Exp+iid baseline. }
\label{tab:simulation-score-deltas}
\end{table}

The results in Table~\ref{tab:simulation-score-deltas} show the intended behaviour of the four model components. When the data are generated from an exponential Hawkes process with iid GPD marks, the more flexible models are essentially tied with the correctly specified Exp+iid baseline, indicating that they incur little penalty from their additional flexibility when it is not required. When cluster-level mark variation is introduced while keeping the exponential triggering kernel, the hierarchical GPD models improve the held-out predictive score. Conversely, when the Hawkes triggering kernel is generated from the non-exponential mixture while the marks remain iid, the DP-Hawkes models improve over the exponential-kernel models, while the hierarchical mark component provides little additional benefit. Finally, when both the mixture triggering kernel and hierarchical mark variation are present, the full DP-Hawkes plus hierarchical GPD model gives the largest improvement.

In summary, the simulation study shows that the two flexible components of the model behave as intended. The DP-Hawkes kernel improves prediction when the true triggering mechanism is non-exponential, while the hierarchical GPD mark model improves prediction when exceedance magnitudes vary across Hawkes-induced clusters. When the corresponding feature is absent from the data-generating mechanism, the additional flexibility incurs little predictive cost.

\section{Real Data Applications} \label{sec:Applications}

\begin{figure*}[t]
   \centering
   \begin{subfigure}[t]{0.48\textwidth}
     \centering
     \includegraphics[width=1\textwidth]{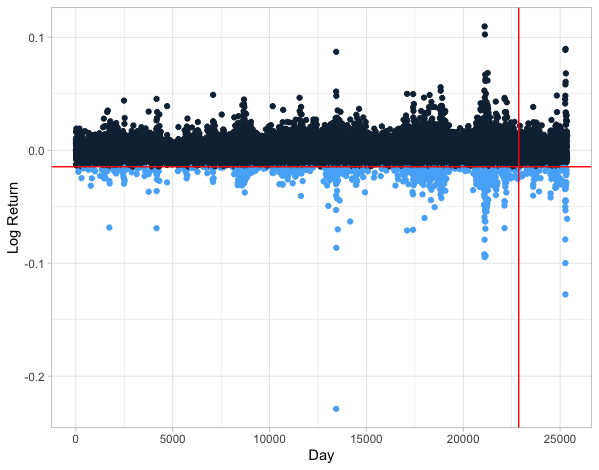}
     \caption{S$\&$P 500, 1951--2020}
     \label{fig:sp500LogReturn}
   \end{subfigure}\hfill
    \begin{subfigure}[t]{0.48\textwidth}
     \centering
     \includegraphics[width=1\textwidth]{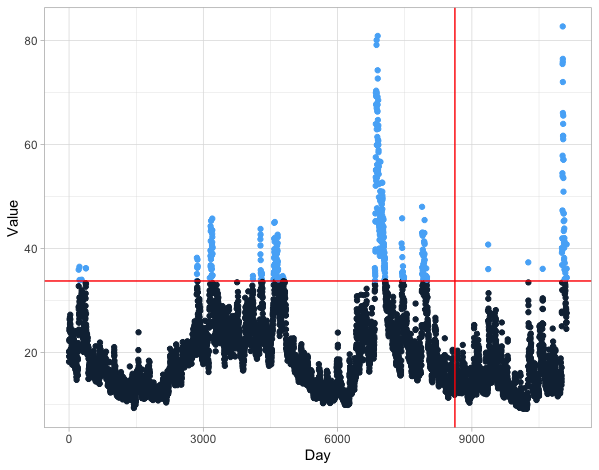}
     \caption{VIX, 1990--2020}
     \label{fig:vixLogreturn}
   \end{subfigure}
      \begin{subfigure}[t]{0.48\textwidth}
     \centering
     \includegraphics[width=1\textwidth]{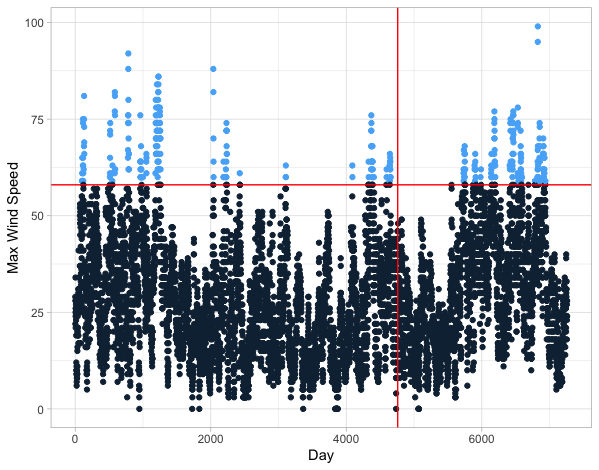}
     \caption{Bradfield wind speeds, 1975--1984}
     \label{fig:Bradfield}
   \end{subfigure}\hfill
    \begin{subfigure}[t]{0.48\textwidth}
     \centering
     \includegraphics[width=1\textwidth]{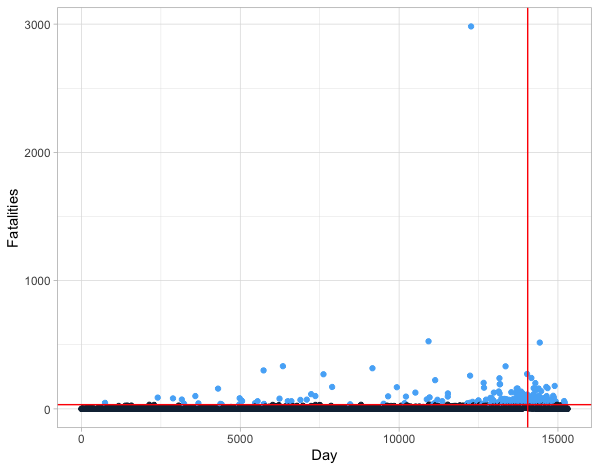}
     \caption{RAND-MIPT Terrorism, 1968--2009}
     \label{fig:logfatalgraph}
   \end{subfigure}

   \caption{Plots of the four real data sets being considered. Light blue points denote the extremes, with the horizontal red line denoting the threshold $u$. The vertical red line shows the separation into training and test sets.}
   \label{fig:realdata}
 \end{figure*}

We next evaluate the model on four real data sets from finance, environmental extremes, and terrorism. The data sets are shown in Figure~\ref{fig:realdata}. In each case the original time series is converted into a sequence of threshold exceedances $(t_i,y_i)$, where $t_i$ is the exceedance time and $y_i=r_i-u$ is the excess above the threshold.

\paragraph{S\&P 500:} The S\&P 500 is a stock market index composed of 500 large companies listed on American stock market exchanges, and is generally thought to be a good representation of the overall U.S. market. Let $x_t$ denote the closing price of the index on day $t$. Since interest is in extreme losses, we define the marks using negative log-returns. The data set consists of daily observations from 1951 to 2020.

\paragraph{VIX:} The VIX (CBOE Volatility Index) is derived from S\&P 500 options and provides a measure of expected market volatility. We obtained daily closing prices of the VIX from 1990 to 2020. Unlike the S\&P 500, the VIX is approximately mean-stationary over a long time horizon, so we define the marks $r_t$ to be the closing price on day $t$ rather than first differences.

\paragraph{Bradfield wind speed:} Extreme wind-speed analysis is a standard application area for Extreme Value Theory \citep{palutikof_review_1999}. We obtained\footnote{\url{http://www.mas.ncl.ac.uk/~nlf8/shortcourse/part3.pdf}} a time series consisting of the maximum hourly observed gust wind speed, in knots, in High Bradfield in the United Kingdom between 1975 and 1984. Here $r_t$ denotes the maximum wind speed during hour $t$.

\paragraph{Terrorism:} The RAND Database of Worldwide Terrorism Incidents (RDWTI) is publicly available\footnote{https://www.rand.org/nsrd/projects/terrorism-incidents/download.html} and contains a detailed list of global terrorism events that occurred between 1968 and 2009. This data set has previously been studied in an extreme-value context by \cite{porter_michael_self-exciting_2012}. For each day during the sample period, we aggregated the total number of worldwide fatalities, so that $r_t$ denotes the number of fatalities on day $t$. The extremely large outlier in Figure~\ref{fig:realdata} corresponds to the 9/11 World Trade Centre attack, while the subsequent increase in terrorism-related fatalities is mostly due to the insurgency in Iraq following the resulting invasion.

\subsection{Evaluation protocol}

For each data set, we choose the extreme-value threshold $u$ to be the 95th percentile of the observed $r_t$ values. In the case of the S\&P 500, where interest lies in extreme losses rather than extreme gains, we use the lower 5th percentile instead. More advanced threshold-selection methods are available \citep{coles_introduction_2001}, but a fixed percentile gives a simple and comparable evaluation across the four applications.

Our proposed model has two main components beyond a standard Hawkes-POT construction: a nonparametric Hawkes triggering kernel and a hierarchical cluster-scale GPD mark model. We therefore compare the same four models as in the simulation study: Exponential or DP-Hawkes
kernels, crossed with either a common iid GPD mark distribution or the
hierarchical cluster-scale GPD mark model.

Performance is assessed using held-out predictive log scores. Each data set is divided into a training set $Y$ and a test set $\tilde{Y}$; the models are estimated on the training set and used to predict both the times and magnitudes of the test-set exceedances. The Bayesian predictive likelihood is
\[
p(\tilde{Y}\mid Y)=\int p(\tilde{Y}\mid\Theta)p(\Theta\mid Y)d\Theta
\approx \frac{1}{M}\sum_{s=1}^M p(\tilde{Y}\mid\Theta^{(s)}),
\]
where $\Theta^{(s)}$ are posterior samples obtained from the training set. For the time component, the held-out point-process likelihood was evaluated sequentially over the observed test events. Each test event was scored conditional on the full training history and all earlier observed test events in the test window. Thus, after a test event occurs, it is allowed to affect the conditional intensity assigned to later test events, just as it would in the likelihood for an observed Hawkes process path. The likelihood also includes the probability of observing no additional events between successive observed test events and after the final test event in the test window. For all data sets except Terrorism, the most recent 10 years are used as the test set. For Terrorism, most extreme events occur in the final decade, so we instead use the most recent five years as the test set to retain sufficient training data for parameter estimation. For numerical stability, the GPD model was fitted to excesses divided by the median positive training excess, with the mean positive training excess used as a fallback if needed. Held-out mark log scores were transformed back to the original excess scale by including the corresponding Jacobian correction.

\begin{figure*}[t]
   \centering
   \includegraphics[width=0.95\textwidth]{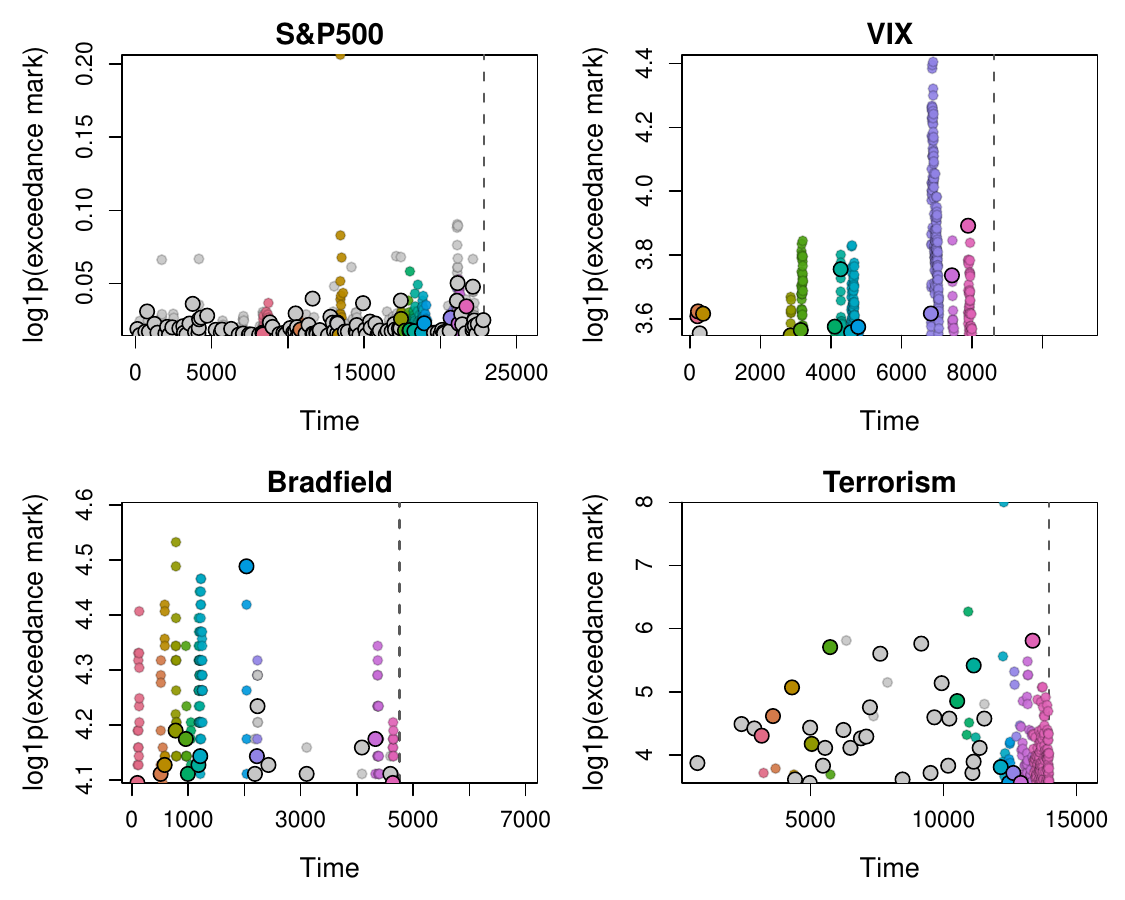}
   \caption{Representative posterior DP-Hawkes clustering draws for the four data sets. Colours distinguish clusters within the displayed posterior draw only; cluster labels are not intended to be matched across MCMC draws. The vertical axis uses a $\log(1+y)$ scale for readability, and the vertical dashed line shows the split between training and test periods.}
   \label{fig:clusters}
\end{figure*}

Figure~\ref{fig:clusters} shows a representative posterior clustering under the DP-Hawkes model for each of the four data sets. These plots are intended as illustrations of the local temporal regimes induced by the Hawkes branching structure, rather than as unique cluster assignments. The induced clustering is then used by the mark model to allow the GPD scale to vary across regimes.

\subsection{Predictive model comparison}

Table~\ref{tab:main-combined-scores} gives the held-out combined log predictive scores for the four model combinations. Higher values indicate better predictive performance. The nonparametric DP-Hawkes kernel improves on the Exponential kernel in every data set when paired with the hierarchical GPD mark model. The full DP-Hawkes plus hierarchical GPD model gives the best score in all four data sets, although the improvement over the DP-Hawkes plus iid GPD model is very small for Terrorism.

\begin{table}[t]
\centering
\begin{tabular}{lrrrr}
\hline
Dataset & Exp+iid & DP+iid & Exp+hier. & DP+hier. \\
\hline
S\&P 500 & -46.62 & -41.84 & -36.42 & \textbf{-33.47} \\
VIX & -312.25 & -284.16 & -301.03 & \textbf{-280.30} \\
Bradfield & -730.14 & -672.77 & -714.76 & \textbf{-659.00} \\
Terrorism & -1525.40 & -1519.45 & -1525.43 & \textbf{-1519.18} \\
\hline
\end{tabular}
\caption{Held-out combined log predictive scores on the original excess scale for the four real data sets. Higher is better; the best score in each row is bolded. Here Exp denotes the parametric Exponential Hawkes kernel, DP denotes the nonparametric DP mixture Hawkes kernel, iid denotes the common GPD mark model, and hier. denotes the hierarchical cluster-scale GPD mark model.}
\label{tab:main-combined-scores}
\end{table}

The comparison in Table~\ref{tab:main-combined-scores} separates the two sources of improvement. Replacing the Exponential Hawkes kernel by the DP mixture kernel generally improves the prediction of exceedance times, while replacing the iid GPD mark model by the hierarchical cluster-scale GPD improves the prediction of exceedance magnitudes. The latter effect is isolated in Table~\ref{tab:dp-mark-scores}, which compares the iid and hierarchical GPD mark models conditional on the DP-Hawkes time model. The hierarchical GPD improves the mark score in all four data sets, with substantial gains for S\&P 500, VIX and Bradfield, and a small gain for Terrorism.

\begin{table}[t]
\centering
\begin{tabular}{lrrr}
\hline
Dataset & iid GPD mark score & hierarchical GPD mark score & improvement \\
\hline
S\&P 500 & 335.07 & 344.15 & \textbf{9.08} \\
VIX & -192.16 & -181.28 & \textbf{10.87} \\
Bradfield & -417.75 & -402.61 & \textbf{15.14} \\
Terrorism & -981.68 & -981.27 & \textbf{0.41} \\
\hline
\end{tabular}
\caption{DP-Hawkes mark-only comparison on the original excess scale. The improvement column is the hierarchical GPD score minus the iid GPD score, and isolates the contribution of the hierarchical mark model conditional on the DP-Hawkes time model. Higher is better.}
\label{tab:dp-mark-scores}
\end{table}

\subsection{Posterior predictive behaviour and fitted mark parameters}

A predictive likelihood can sometimes hide poor tail behaviour. We therefore also examine posterior predictive summaries for the maximum held-out excess under the full DP-Hawkes plus hierarchical GPD model. Table~\ref{tab:predictive-maxima-summary} compares the observed held-out maximum with the posterior predictive median and central 90\% predictive interval. The observed maxima for S\&P 500, VIX and Bradfield are broadly consistent with the posterior predictive distributions. The terrorism data set has a much heavier fitted tail, leading to a wider predictive distribution for the maximum future excess; this is consistent with the posterior estimate of the positive GPD shape parameter for that data set.

\begin{table}[t]
\centering
\begin{tabular}{lrrr}
\hline
Dataset & observed maximum & predictive median & 90\% predictive interval \\
\hline
S\&P 500 & 0.113 & 0.045 & (0.019, 0.124) \\
VIX & 48.95 & 14.45 & (1.52, 59.80) \\
Bradfield & 41.00 & 34.27 & (14.94, 71.05) \\
Terrorism & 484.0 & 634.7 & (194.0, 2464) \\
\hline
\end{tabular}
\caption{Posterior predictive summaries for the maximum held-out excess under the full DP-Hawkes plus hierarchical GPD model. Predictive intervals are central 90\% intervals, and all values are reported on the original excess scale.}
\label{tab:predictive-maxima-summary}
\end{table}

\begin{figure*}[t]
   \centering
   \includegraphics[width=0.90\textwidth]{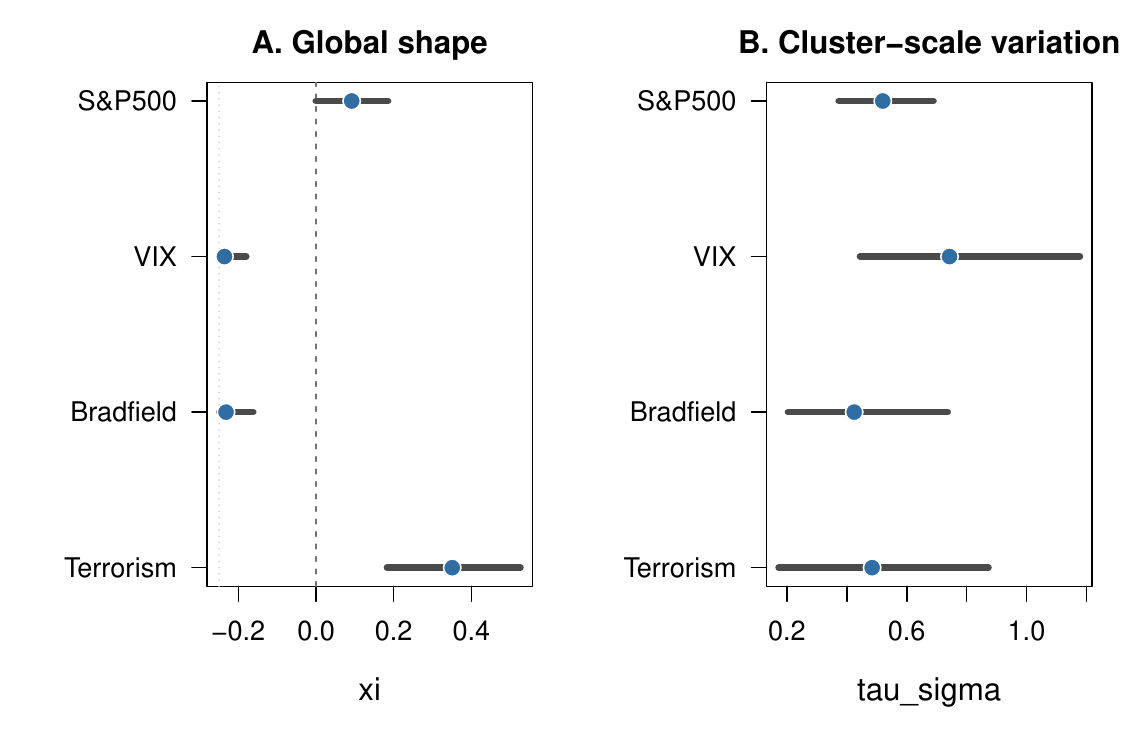}
   \caption{Posterior medians and 95\% credible intervals for the main GPD mark parameters under the full DP-Hawkes plus hierarchical GPD model. Panel A shows the shared GPD shape parameter $\xi$, with reference lines at $\xi=0$ and at the lower truncation point $\xi=-0.25$. Panel B shows the cluster-scale variation parameter $\tau_\sigma$.}
   \label{fig:gpd-parameters}
\end{figure*}

Figure~\ref{fig:gpd-parameters} summarises the posterior distribution of the main GPD mark parameters under the full DP-Hawkes model. The global shape parameter $\xi$ varies substantially between applications: it is positive for Terrorism, mildly positive for S\&P 500, and close to the lower regularisation boundary for VIX and Bradfield. The posterior for $\tau_\sigma$ is away from zero in all four data sets, indicating that the Hawkes-induced clusters contain useful information about variation in the scale of exceedance magnitudes.

\section{Discussion}

In this paper we have developed and applied a novel framework for modelling
extreme events that relaxes some of the conditions of standard extreme value
theory and allows for nonstationarity in both the exceedance times and
magnitudes. Our key focus is on the predictive modelling of future extreme
values, which distinguishes our work from previous nonparametric Bayesian point
process models which are essentially based on retrospective smoothing of
historical extremes \citep{kottas_bayesian_2007,kottas_spatial_2012}. We also
developed a posterior simulation algorithm for the resulting model, and showed
through both simulation and real-data examples that predictive accuracy can be
improved by taking both types of nonstationarity into account.

The simulation and empirical results support both components of the proposed model. In the real-data applications, replacing the parametric Exponential Hawkes
kernel with a nonparametric Dirichlet process mixture kernel improved held-out
predictive performance across the four data sets when paired with the
hierarchical mark model. The simulation study shows the corresponding
controlled behaviour: the DP kernel improves prediction when the true
triggering pattern is non-exponential. This suggests that the temporal clustering of
extreme events is not always adequately captured by a simple parametric
excitation kernel. Second, conditional on the DP-Hawkes time model, replacing
the iid GPD mark distribution with the hierarchical cluster-scale GPD improved
the mark score in all four applications. The improvement was substantial for
S\&P 500, VIX and Bradfield wind speeds, and smaller for the terrorism data,
where the iid and hierarchical mark models gave similar predictive performance.
Taken together, these results indicate that Hawkes-induced clustering is useful
not only for predicting when extremes occur, but also for modelling changes in
their magnitudes. Together, the simulation and real-data results indicate that the two flexible
components are useful for different aspects of nonstationarity.

A key modelling choice in the mark distribution is to allow the GPD scale to
vary across Hawkes-induced clusters while estimating the shape parameter
globally. This gives a compromise between a stationary POT model, which treats
all exceedance magnitudes as iid draws from a single GPD, and a fully local
model which attempts to estimate separate tail parameters within each cluster.
The latter is unattractive in this setting because the branching structure can
produce many small clusters, making cluster-specific tail-shape estimation
unstable. The shared-shape specification retains the ability to capture local
changes in the typical magnitude of exceedances while borrowing information
across the full data set for tail-shape inference. The posterior summaries in the
real-data examples show that the fitted shape parameter varies substantially
between applications, while the cluster-scale variation parameter is away from
zero in each case.

Several extensions would be worth pursuing. The background intensity
\(\mu(t)\) could be made time-varying or allowed to depend on covariates,
allowing long-run changes in the rate of exceedances to be separated from
short-run self-excitation. Similarly, covariates could be introduced into the
mark distribution, either through the cluster-scale hierarchy or through the
global shape parameter. A multivariate extension would also be valuable in
applications such as financial risk or natural hazards, where extremes in
different series or regions may interact. More generally, the results suggest
that combining self-exciting point process models with hierarchical extreme
value models is a promising approach for forecasting nonstationary extremes.

\section*{Statements and Declarations}

\subsection*{Competing interests}
The authors declare that they have no competing interests.

\subsection*{Funding}
No funding was received for this work.

\subsection*{Data availability}
The data sets analysed in this article are publicly available from the sources
described in Section 6.

\subsection*{Code availability}
Code implementing the methods and reproducing the analyses is available from
the corresponding author upon reasonable request.

\clearpage

\appendix
\setcounter{figure}{0}
\renewcommand{\thefigure}{A.\arabic{figure}}

\section{MCMC diagnostics}
\label{app:mcmc}

\begin{figure}[t]
\centering
\includegraphics[width=\textwidth]{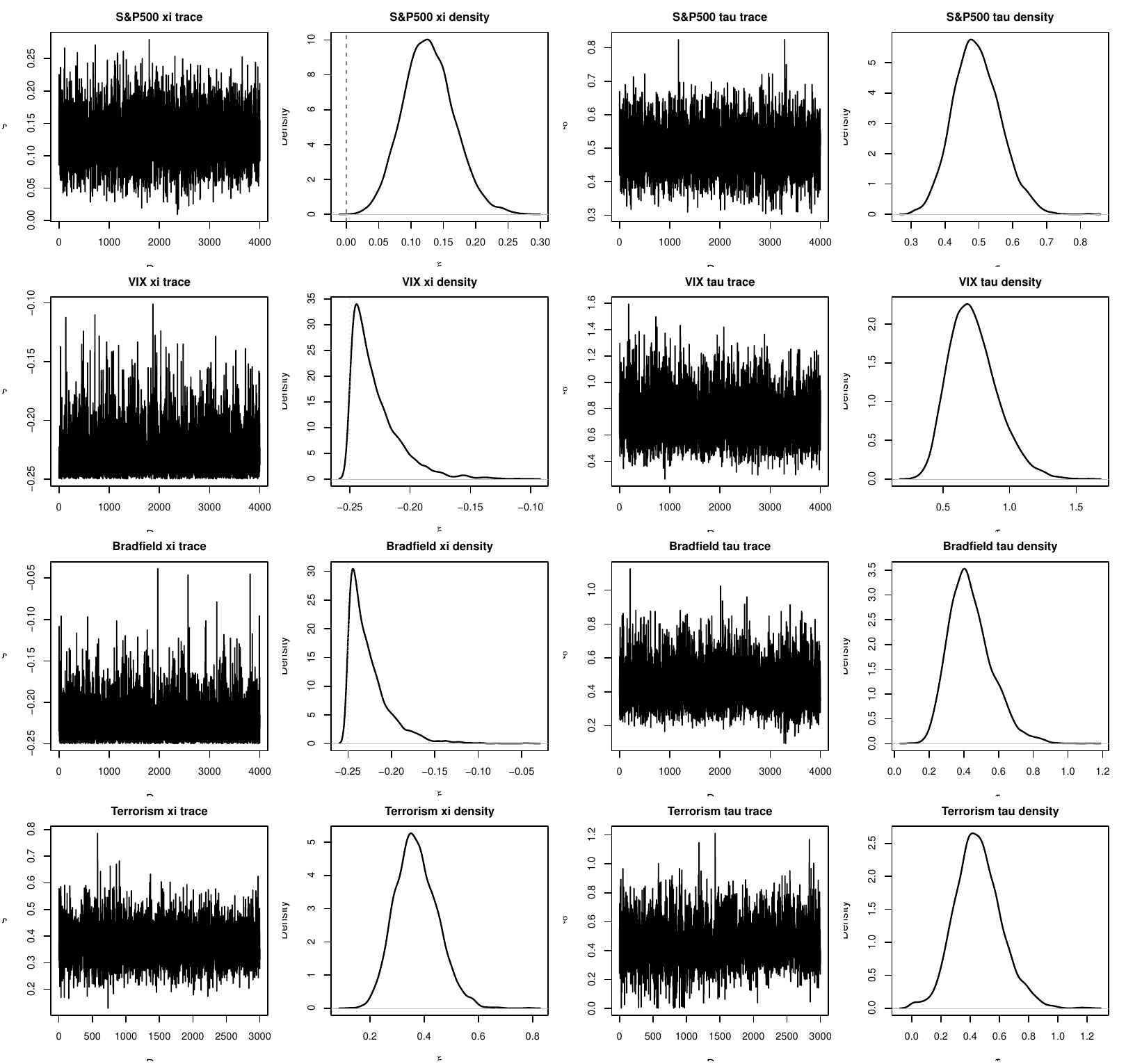}
\caption{Trace and posterior density plots for \(\xi\) and \(\tau_\sigma\) under
the final DP-Hawkes plus hierarchical GPD model.}
\label{fig:mcmc-traces}
\end{figure}

Figure~\ref{fig:mcmc-traces} shows trace and posterior density plots for the
global GPD shape parameter \(\xi\) and the cluster-scale variation parameter
\(\tau_\sigma\) under the final DP-Hawkes plus hierarchical GPD model. These
diagnostics provide an additional check on the final real-data fits.

\clearpage
\pagebreak

\bibliography{zotero}

\end{document}